\documentclass[aps,prd,onecolumn,groupedaddress,showpacs,nofootinbib,amssymb]{revtex4}
\usepackage[dvips]{graphicx}
\usepackage{amssymb}
\usepackage{amsmath}
\usepackage{graphicx}
\usepackage{amsfonts}
\usepackage{bm}
\usepackage{latexsym,float}
\usepackage{epsfig}
\usepackage{epstopdf}
\usepackage{amssymb}
\usepackage{verbatim}
\usepackage{multirow}
\usepackage{color}

\begin{document}

\title{Deceleration versus Acceleration Universe in Different Frames of $F(R)$ Gravity}

\author{Sebastian Bahamonde$^{1}$\,\thanks{sebastian.beltran.14@ucl.ac.uk},
Sergei~D.~Odintsov$^{2,3}$\,\thanks{odintsov@ieec.uab.es}, V.K.
Oikonomou$^{4,5}$\,\thanks{v.k.oikonomou1979@gmail.com},Petr V.
Tretyakov$^{6,7}$\,\thanks{tpv@theor.jinr.ru}} \affiliation{$^{1)}$
Department of Mathematics, University College London,
    Gower Street, London, WC1E 6BT, United Kingdom \\
$^{2)}$Institut de Ciencies de lEspai (IEEC-CSIC),
    Campus UAB, Carrer de Can Magrans, s/n, 08193 Cerdanyola del Valles, Barcelona, Spain \\
$^{3)}$ICREA, Passeig LluAs Companys, 23,
    08010 Barcelona, Spain\\
$^{4)}$  Tomsk State
Pedagogical University, 634061 Tomsk, Russia\\
$^{5)}$Laboratory for Theoretical Cosmology, Tomsk State University
of Control Systems and Radioelectronics (TUSUR), 634050 Tomsk,
Russia\\
$^{6)}$Joint Institute for Nuclear Research, Joliot-Curie 6, 141980
Dubna, Moscow region, Russia\\
$^{7)}$Institute of Physics, Kazan Federal University, Kremlevskaya
street 18, 420008 Kazan, Russia}

\begin{abstract}
In this paper we study the occurrence of accelerating universe
versus decelerating universe between the F(R) gravity frame (Jordan
frame) and non-minimally coupled scalar field theory frame, and the
minimally coupled scalar field theory frame (Einstein frame) for
various models. As we show, if acceleration is imposed in one frame,
it will not necessarily correspond to an accelerating metric when
transformed in another frame. As we will demonstrate, this issue is
model and frame-dependent but it seems there is no general scheme
which permits to classify such cases.
\end{abstract}

\pacs{04.50.Kd, 95.36.+x, 98.80.-k, 98.80.Cq,11.25.-w}

\maketitle

\section{Introduction}

One of the most profound questions in modified gravity is related
with the choice of the physical frame. The frame corresponding to
$F(R)$ gravity \cite{reviews1} is called Jordan frame, and by a
conformal transformation it can be transformed to a minimally
coupled scalar-tensor theory, with the corresponding frame being
called Einstein frame. In addition to these, there are also frames
in which the scalar field is non-minimally coupled to gravity, and
these can be reached by a $F(R)$ gravity by also using a suitably
chosen conformal transformation, or directly by the Einstein frame
theory by conformally transforming the theory.

Generally speaking, one should confront the theoretical predictions
of a specific gravitational theory with the observable Universe
history supported by the current observational data. In this sense,
each of the three mentioned frames, namely the $F(R)$ gravity, and
the minimal and non-minimal scalar theories, may give a viable
description of the observable Universe history. However, it is not
sure that a viable description in one frame gives also viable and
convenient description in the other frame. For instance, it may give
a viable but physically inconvenient description. In other words,
there appears the question which of these three frames is the most
physical one (and in which sense) or, at least, which of these
frames gives a convenient description of the Universe history.
Eventually, the answer to this question depends very much from the
confrontation with the observational data, from the specific choice
of the theory and from the observer associated with specific frame.
At the same time, the related question is about equivalent results
in all three frames and/or about construction of the observable
quantities which are invariant under conformal transformations
between the three frames.

In the study of the inflationary epoch, when one is dealing with
quasi-de Sitter space, it is expected that the spectral index of
primordial curvature perturbations and the scalar-to-tensor ratio
calculated in two frames ($F(R)$ and minimal scalar-tensor) are
nearly the same. Indeed, the equivalence of two frames was
explicitly demonstrated in Refs. \cite{kaizer} and also in
\cite{Brooker:2016oqa}. However, this is surely not enough for
number of reasons. For example, for the e?ects on neutron stars in
$F(R)$ gravity, the Jordan and Einstein frame pictures are di?erent,
as was shown in Ref. \cite{capp}.

 In addition, finite-time singularities
\cite{Nojiri:2005sx,Barrow:2004xh,noo4,noo5,noo6} between Jordan and
Einstein frames belong to different types of singularity, see for
example \cite{noo5,bahamonte}, because the conformal transformation
does not work for singular points. In this research line, in this
paper we shall investigate under which circumstances, an
accelerating evolution in one frame may be transformed to a
decelerating evolution in the other frame. We shall consider three
types of frames, namely the $F(R)$ gravity frame (Jordan frame), the
minimally coupled scalar-tensor theory frame (Einstein frame) and
the non-minimally coupled scalar-tensor frame. By using several
illustrative examples we shall demonstrate that an accelerating
cosmology in one frame may correspond to a decelerating cosmology in
the other frame, and thus the physical interpretation in the two
frames may be different.

For a preliminary discussion along this research line, see
\cite{Capozziello:2006dj}. Eventually, it depends on the model under
investigation and on the specific choice of the conformal
transformation. For simplicity, we do not add matter sector in this
paper, since this may introduce extra complications due to the
appearance of a non-minimal coupling in the matter sector for the
Einstein frame.

This paper is organized as follows: In Sec.~\ref{sec:0} we introduce some essential information about the correspondence between the Jordan and Einstein frame. In Sec.~\ref{sec:1} we study the case where the scalar field is
minimally coupled to the scalar curvature, so we start from the
Einstein frame, and we study how an accelerating cosmology in the
Einstein frame is transformed in the Jordan frame. Sec.~\ref{sec:2}
is devoted to generalize the latter idea by taking into account a
theory non-minimally coupled between the scalar field and the Ricci
scalar, so we study the correspondence between these frames. Finally
the conclusions follow in Sec.~\ref{sec:3}.

\section{Correspondence Between Jordan and Einstein Frame: Essential Properties}\label{sec:0}

Before we go into the main focus of the paper, we present in brief
some essential information regarding the correspondence between the
Jordan and Einstein frame. For details on these issues, we refer the
reader to Ref. \cite{reviews1}. For simplicity, in this section we will assume that $\kappa^2=1$. Let us start from the Jordan frame
$F(R)$ gravity action, 
\begin{equation}\label{pure}
\mathcal{S}=\frac{1}{2}\int\mathrm{d}^4x \sqrt{-\hat{g}}F(R)\, ,
\end{equation}
with $\hat{g}_{\mu \nu}$ being the metric tensor in the Jordan
frame. By introducing an auxiliary field, which we denote as $A$,
the action of Eq. (\ref{pure}) is written in the following way,
\begin{equation}\label{action1dse111}
\mathcal{S}=\frac{1}{2}\int \mathrm{d}^4x\sqrt{-\hat{g}}\left (
F'(A)(R-A)+F(A) \right )\, .
\end{equation}
Upon variation of the action (\ref{action1dse111}), with respect to
the auxiliary scalar degree of freedom, we obtain the solution
$A=R$, and this actually proves that the actions of Eqs.
(\ref{action1dse111}) and (\ref{pure}) are mathematically
equivalent. The Jordan and Einstein frames are connected via the
following canonical transformation,
\begin{equation}\label{can}
\varphi =\sqrt{\frac{3}{2}}\ln (F'(A))
\end{equation}
where $\varphi$  will denote the canonical scalar field in the
Einstein frame. By making the following  conformal transformation,
\begin{equation}\label{conftransmetr}
g_{\mu \nu}=e^{-\varphi }\hat{g}_{\mu \nu }
\end{equation}
with $g_{\mu \nu}$ denoting the Einstein frame metric, we finally
obtain the following action,
\begin{align}\label{einsteinframeaction}
& \mathcal{\tilde{S}}=\int \mathrm{d}^4x\sqrt{-g}\left (
R-\frac{1}{2}\left (\frac{F''(A)}{F'(A)}\right )^2g^{\mu \nu
}\partial_{\mu }A\partial_{\nu }A -\left (
\frac{A}{F'(A)}-\frac{F(A)}{F'(A)^2}\right ) \right ) \\ \notag & =
\int \mathrm{d}^4x\sqrt{-g}\left ( R-\frac{1}{2}g^{\mu \nu
}\partial_{\mu }\varphi\partial_{\nu }\varphi -V(\varphi )\right )\,
,
\end{align}
which is the Einstein frame action of the canonical scalar field
$\varphi $. The scalar field potential $V(\varphi )$ appearing in
Eq. (\ref{einsteinframeaction}), is equal to,
\begin{align}\label{potentialvsigma}
V(\varphi
)=\frac{1}{2}\left(\frac{A}{F'(A)}-\frac{F(A)}{F'(A)^2}\right)=\frac{1}{2}\left
( e^{-\sqrt{2/3}\varphi }R\left (e^{\sqrt{2/3}\varphi} \right )-
e^{-2\sqrt{2/3}\varphi }F\left [ R\left (e^{\sqrt{2/3}\varphi}
\right ) \right ]\right )\, .
\end{align}
Therefore, for a specifically given $F(R)$ gravity, we can find the
corresponding canonical scalar field potential by using the
expression (\ref{potentialvsigma}). The method can work in the
converse way, by finding the function $R(\varphi)$. We can easily
express the Ricci scalar as a function of the canonical scalar
field, by solving Eq. (\ref{can}) with respect to the auxiliary
scalar $A$, bearing in mind that $A=R$. Then, for a specifically
given potential, we can combine Eqs. (\ref{potentialvsigma}) and
(\ref{can}), and by differentiating with respect to $R$, we obtain,
\begin{equation}\label{solvequation}
RF_R=2\sqrt{\frac{3}{2}}\frac{\mathrm{d}}{\mathrm{d}\varphi}\left(\frac{V(\varphi)}{e^{-2\left(\sqrt{2/3}\right)\varphi}}\right)
\end{equation}
where $F_R=\frac{\mathrm{d}F(R)}{\mathrm{d}R}$. In this way, if the
scalar potential is given, the $F(R)$ gravity easily follows by
using Eq. (\ref{solvequation}).

\section{MINIMALLY CURVATURE-COUPLED SCALAR-TENSOR THEORY}\label{sec:1}

Let us start from the minimally coupled scalar-tensor theory action,
which is the Einstein frame action,
\begin{equation}
S=\int d^4x \sqrt{-g} \left\{ \frac{1}{2\kappa^2}R -
\frac{1}{2}\partial_{\mu}\phi\partial^{\mu}\phi - V(\phi) \right\}.
\label{1.1}
\end{equation}
By assuming a flat Friedmann-Robertson Walker (FRW) metric in the
Einstein frame, with line element,
\begin{equation}
ds^2_{E}=  -dt^2 + a(t)^2 d\mathbf{x}^2, \label{1.1.1}
\end{equation}
the equations of motion corresponding to the action (\ref{1.1}) are
equal to,
\begin{equation}
3H^2=\frac{1}{2}\dot\phi^2 +V, \label{1.2}
\end{equation}
\begin{equation}
3H^2+2\dot H=-\frac{1}{2}\dot\phi^2 +V, \label{1.3}
\end{equation}
where for simplicity we used a physical units system where
$\kappa^2=1$.

Now we conformally transform the metric $g_{\mu\nu}=
e^{\sqrt{\frac{2}{3}} \phi} \tilde{g}_{\mu\nu}$ to switch the
Einstein frame to the Jordan frame, in which the action of Eq.
(\ref{1.1}) takes the form 
\begin{equation}
S=\int d^4x \sqrt{-\tilde{g}} \left\{
\frac{1}{2}e^{\sqrt{\frac{2}{3}} \phi}\tilde{R}  -
e^{2\sqrt{\frac{2}{3}} \phi}  V(\phi) \right\}. \label{1.4}
\end{equation}
Hence the line element of the Jordan frame metric reads,
\begin{equation}
ds^2_{J}= e^{\sqrt{\frac{2}{3}} \phi} \left ( -dt^2 + a(t)^2
d\mathbf{x}^2  \right ), \label{1.5}
\end{equation}
or by introducing a new time coordinate $\tilde{t}$, which is
defined as follows,
\begin{equation}
d\tilde{t}=e^{\frac{1}{2}\sqrt{\frac{2}{3}} \phi} dt, \label{1.6}
\end{equation}
the metric of Eq. (\ref{1.5}) may be rewritten as follows,
\begin{equation}
ds^2_{J}= -d\tilde{t}^2 + \tilde{a}(\tilde{t})^2 d\mathbf{x}^2,
\label{1.7}
\end{equation}
and the scale factors of the Jordan and Einstein frames are related
as follows,
\begin{equation}
\tilde{a}(\tilde{t}(t))=e^{\frac{1}{2}\sqrt{\frac{2}{3}} \phi} a(t).
\label{1.8}
\end{equation}

For an arbitrary function $b$ wich depends on time, we will use the following notation: $\dot{b}\equiv \frac{db}{dt}$ and $b'\equiv \frac{db}{d\tilde{t}}$. Now, if we calculate first derivative of expression (\ref{1.8}):
\begin{equation}
\tilde{a}'=\frac{1}{2}\sqrt{\frac{2}{3}} \dot\phi a + \dot a,
\label{1.9}
\end{equation}
and also the second derivative reads,
\begin{equation}
\tilde{a}''=\left (\frac{1}{2}\sqrt{\frac{2}{3}} \ddot\phi a +
\frac{1}{2}\sqrt{\frac{2}{3}} \dot\phi \dot a+ \ddot a \right
)e^{-\frac{1}{2}\sqrt{\frac{2}{3}} \phi}. \label{1.10}
\end{equation}
Now the conditions that must be satisfied in order for an
accelerated expansion in one frame corresponds to decelerated
expansion in another one, are $\ddot a>0$ and simultaneously
$\tilde{a}''<0$. In addition the conditions $\dot a>$ and
$\tilde{a}'>0$ must hold true in order to have accelerating
expansion in the both frames. It is clear from expression
(\ref{1.10}) that in order for the above constraints to be
satisfied, it suffices if the following conditions hold true,
\begin{equation}
A\equiv\frac{1}{2}\sqrt{\frac{2}{3}} \ddot\phi a +
\frac{1}{2}\sqrt{\frac{2}{3}} \dot\phi \dot a+ \ddot a >0,
\label{1.11}
\end{equation}
\begin{equation}
\ddot a <0, \label{1.12}
\end{equation}
and $\dot a>$ and $\tilde{a}'>0$ as well.

Consider the following cosmological evolution in the Einstein frame,
\begin{equation}
H(t)\equiv\frac{\dot a}{a}=f_0(t-t_s)^{\alpha}, \label{1.13}
\end{equation}
where for simplicity we will assume that $t_s=0$. By integrating Eq.
(\ref{1.13}) we find,
\begin{equation}
 a(t)=a_0 e^{\frac{f_0}{\alpha+1}t^{\alpha+1}}.
\label{1.14}
\end{equation}
We can see that expressions for $\tilde{a}'$ and $\tilde{a}''$
contain only $a(t)$, $\phi(t)$ and its derivatives with respect to
the cosmic time $t$. In order to find $\phi(t)$ let us subtract
equations (\ref{1.2})-(\ref{1.3}), and we find,
\begin{equation}
-2\dot H=\dot\phi^2, \label{1.15}
\end{equation}
which is true for any type of potential $V(\phi)$. Therefore, by
using Eq. (\ref{1.13}) we obtain,
\begin{equation}
\dot\phi(t)=\sqrt{-2f_0\alpha}t^{\frac{\alpha-1}{2}}, \label{1.16}
\end{equation}
and in addition we get,
\begin{equation}
\ddot\phi(t)=\frac{\alpha-1}{2}\sqrt{-2f_0\alpha}t^{\frac{\alpha-3}{2}}.
\label{1.17}
\end{equation}
First of all, note that we need a positive parameter $f_0$ for an
expanding Universe in the Einstein frame, and in addition, the
parameter $\alpha$ should be negative in order to have real values
of the scalar field. The first derivative $\dot a$ reads,
\begin{equation}
\dot a=af_0t^{\alpha}, \label{1.18}
\end{equation}
and it is always positive for the parameters chosen as we discussed
above. By taking into account Eq. (\ref{1.9}), the first derivative
$\tilde{a}'$ reads,
\begin{equation}
\tilde{a}'=af_0t^{\alpha}\left( 1 + \sqrt{\frac{-\alpha}{3f_0}}
t^{\frac{-\alpha-1}{2}} \right ), \label{1.19}
\end{equation}
and we can see that for any set of parameters, it is impossible to
have expansion in one frame and contraction in the other frame. Now
let us calculate the second derivatives of the scale factors. For
$\ddot a$ we have,
\begin{equation}
\ddot a=\dot af_0t^{\alpha}+af_0\alpha
t^{\alpha-1}=af_0t^{\alpha-1}\left( \alpha + f_0 t^{\alpha+1} \right
), \label{1.20}
\end{equation}
and we can see that depending on time, this function may have
different sign for the parameters chosen as above. By calculating
the expression (\ref{1.11}), we get,
\begin{equation}
A= a t^{\frac{\alpha-3}{2}} \left(
\frac{\alpha-1}{2}\sqrt{\frac{-f_0\alpha}{3}}
+\sqrt{\frac{-f_0\alpha}{3}}f_0 t^{\alpha+1}
+f_0^2t^{\frac{3\alpha+3}{2}} + f_0\alpha t^{\frac{\alpha+1}{2}}
\right ). \label{1.21}
\end{equation}
Thus, according to our previous considerations, we are interested in
the case that $\ddot a<0$ and $A>0$ or equivalently,
\begin{equation}
\alpha + f_0 t^{\alpha+1}<0, \label{1.22}
\end{equation}
and in addition,
\begin{equation}
\frac{\alpha-1}{2}\sqrt{\frac{-f_0\alpha}{3}}
+\sqrt{\frac{-f_0\alpha}{3}}f_0 t^{\alpha+1}
+f_0^2t^{\frac{3\alpha+3}{2}} + f_0\alpha t^{\frac{\alpha+1}{2}}>0.
\label{1.23}
\end{equation}
From a general point of view we have the next situation: two
inequalities (\ref{1.22})-(\ref{1.23}) with three parameters
$\alpha$, $f_0$ and $t$. So it is quite natural to expect that both
these inequalities may be satisfied at least for some time instance.
But actually the real picture is more complicated, because there are
additional restrictions for the parameters, namely, $\alpha<0$,
$f_0>0$, which must be satisfied as well. The inequality of Eq.
(\ref{1.22}) indicates,
\begin{equation}
t^{\alpha+1}<\frac{-\alpha}{f_0}. \label{1.24}
\end{equation}
Let us suppose that (\ref{1.24}) may hold true at the time instance
$t_*$, thus we may put,
\begin{equation}
t^{\alpha+1}_*=m\frac{-\alpha}{f_0}, \label{1.25}
\end{equation}
where $m$ is some numerical parameter restricted according to
(\ref{1.24}) as
\begin{equation}
0<m<1. \label{1.26}
\end{equation}
Substituting expression (\ref{1.25}) in  (\ref{1.23}), we find,
\begin{equation}
\frac{\alpha-1}{2}\sqrt{\frac{-f_0\alpha}{3}} -\alpha
m\sqrt{\frac{-f_0\alpha}{3}} +
m^{\frac{3}{2}}|\alpha|\sqrt{-f_0\alpha} +
m^{\frac{1}{2}}\alpha\sqrt{-f_0\alpha}>0. \label{1.27}
\end{equation}
Note at this point, that in expression (\ref{1.23}) the first and
fourth terms are negative, whereas the second and the third are
positive. Also note that in Eq. (\ref{1.27}) we have $\alpha<0$. The
above expression can be simplified as follows,
\begin{equation}
B=\frac{\alpha-1}{2\sqrt{3}} -\frac{\alpha m}{\sqrt{3}} +
m^{\frac{3}{2}}|\alpha| + m^{\frac{1}{2}}\alpha>0. \label{1.28}
\end{equation}
We rewrite the above expression as follows,
\begin{equation}
A_1+A_2>0, \label{1.29}
\end{equation}
where $A_1$ is equal to,
\begin{equation}
A_1=\frac{\alpha-1}{2\sqrt{3}} -\frac{\alpha m}{\sqrt{3}},
\label{1.30}
\end{equation}
while $A_2$ is,
\begin{equation}
A_2=m^{\frac{3}{2}}|\alpha| + m^{\frac{1}{2}}\alpha. \label{1.31}
\end{equation}
We can see that $A_2$ is always negative in the range $(\alpha<0$,
$0<m<1)$, whereas $A_1$ may be positive due to the last term. Thus
it is clear Eq. (\ref{1.28}) will hold true if $A_1>0$. A detailed
analysis of this inequality, imposes additional restrictions for
parameters, namely, $\alpha<-1$ and $\frac{1}{2}<m<1$. By taking
into account the negative values of $\alpha$, it is possible to
rewrite (\ref{1.28}) as follows,
\begin{equation}
|\alpha|>
\frac{1}{2\sqrt{3}m^{\frac{3}{2}}+2m-2\sqrt{3}m^{\frac{1}{2}}-1}.
\label{1.32}
\end{equation}
The expression in the denominator is monotonically increasing
function of $m$ in the range $\frac{1}{2}<m<1$, which crosses zero
near the point $m\thickapprox 0.8042$. This means, that all
interesting for us values of $m$ lie in the narrow interval
$0.8042\lesssim m<1$. Let us take for instance $m=0.9$, then
expression (\ref{1.32}) indicates that $|\alpha|\gtrsim 2.12$, so
let us assume $\alpha=-3$. By substituting these values into
(\ref{1.28}) we find $B\backsimeq 0.119$. In addition, note that the
result is true for any positive values of $f_0$, whereas the time
$t$ may be calculated by using (\ref{1.25}). Thus we explicitly
demonstrated how to obtain an accelerating cosmology in one frame,
which corresponds to a decelerating one in a conformal frame, for
the Jordan and Einstein frames. In the next section we consider the
non-minimally coupled scalar field frame.

\section{NON-MINIMALLY CURVATURE-COUPLED SCALAR THEORY}\label{sec:2}

Let us now consider the case where the Ricci scalar is non-minimally
coupled to the scalar field. In this case, the gravitational action
takes the following form,
\begin{equation}
S=\int d^{4}x\sqrt{-g} \left[(1+f(\phi))\frac{R}
{\kappa^{2}}-\frac{1}{2} \omega (\phi ) \partial_{\mu }\phi
\partial^{\mu }\phi -V(\phi )\right]\,.
\label{2.1}
\end{equation}
We shall refer to this non-minimally coupled frame as Jordan frame
too. In Ref. \cite{Elizalde:2008yf}, it was shown that by performing
the following conformal transformation,
\begin{equation}
g_{\mu \nu }=[1+f(\phi )]^{-1} \tilde{g}_{\mu \nu }\ , \label{2.2}
\end{equation}
we recover the Einstein frame minimally-coupled scalar-tensor theory
given by the action,
\begin{equation}
S=\int dx^{4} \sqrt{-\tilde{g}}\left( \frac{\tilde{R}}{ \kappa
^{2}}- \frac{1}{2} \, W( \phi ) \, \tilde{g}^{\mu\nu}\partial_{\mu
}\phi \partial_{\nu} \phi -U(\phi)\right)\ , \label{2.4}
\end{equation}
where the functions $W(\phi)$ and $U(\phi)$ are defined as follows,
\begin{eqnarray}
W(\phi )&=&\frac{ \omega
    (\phi )}{1+f(\phi)}
+\frac{3}{\kappa ^{2}(1+f(\phi ))^{2}}
\left( \frac{df(\phi )}{d\phi }\right) ^{2}\,,\label{W}\\
U(\phi )&=&\frac{ V(\phi )}{ \left[ 1+f(\phi )
\right]^2}\,.\label{U}
\end{eqnarray}
For the action of Eq. (\ref{2.4}) and for a flat FRW metric, the
cosmological equations can be written as follows (taking $\kappa^2=1$),
\begin{eqnarray}
    && \tilde{H}^{2}=\frac{\kappa ^{2}}{6}\rho _{\phi } \;,\label{2.5}\\
    &&\nonumber \\
    && \tilde{H}'=-\frac{\kappa ^{2}}{{4}}\left( \rho
    _{\phi } + p_{\phi}\right)\ , \label{2.6}\\
    &&\nonumber \\
    &&  2W(\phi)\left[\phi'' +3\tilde{H} \,
    \phi'\right] +\left[ W_{\phi}
    \left( \phi'\right)^2 +2U_{\phi} \right] =0
    \;,
\end{eqnarray}
where $\rho_{\phi}$ stands for,
\begin{align}
\rho _{\phi }&=\frac{1}{2}W(\phi ) (\phi')^{2} +
U(\phi )\,,\label{rho}\\
p_{\phi }&=\frac{1}{2}W(\phi )(\phi')^{2} - U(\phi )\,,\label{p}
\end{align}
and the ``prime'' denotes differentiation with respect to
$\tilde{t}$, whereas $W_{\phi}$ denotes partial differentiation with
respect to $\phi$. Therefore, we have that
\begin{align}
W(\phi )(\phi')^{2} &=-4\tilde{H}'\,,\label{2.7}\\
U(\phi )&=6\tilde{H}^{2}+2\tilde{H}'. \label{2.8}
\end{align}
In the rest of this section, we will study how the acceleration
might change from one frame and another.

\subsection{How does acceleration change from one frame to another?}

Let us now find the conditions which when are satisfied, we can have
acceleration in one frame and deceleration in the other. The scale
factors and time-coordinates are related by,
\begin{align}
a(t)=[1+f(\phi)]^{-1/2} \, \tilde{a}(\tilde{t})\,, \ \
\frac{d\tilde{t}}{dt}= [1+f(\phi)]^{1/2}\ . \label{2.11}
\end{align}
Now, by differentiating with respect to the time $t$, we find,
\begin{align}
\frac{d a(t)}{d t}\equiv \dot
a=\tilde{a}'-\frac{1}{2}[1+f(\phi)]^{-1}f_{\phi}\phi'\tilde{a},
\label{2.12}
\end{align}
and the second derivative is,
\begin{align}
\ddot a=\frac{1}{2}[1+f(\phi)]^{-\frac{1}{2}} \left\{
2[1+f(\phi)]\tilde{a}'' +
[1+f(\phi)]^{-1}f_{\phi}^2(\phi')^2\tilde{a}
-f_{\phi\phi}(\phi')^2\tilde{a} -
f_{\phi}\phi''\tilde{a}-f_{\phi}\phi'\tilde{a}'  \right\}.
\label{2.13}
\end{align}
Now, let us assume that in the Jordan (Einstein) frame the Universe
is decelerating (accelerating). To materialize such a scenario, the
following inequalities need to hold true,
\begin{eqnarray}
2[1+f(\phi)]\tilde{a}'' + [1+f(\phi)]^{-1}f_{\phi}^2(\phi')^2\tilde{a} -f_{\phi\phi}(\phi')^2\tilde{a} - f_{\phi}\phi''\tilde{a}-f_{\phi}\phi'\tilde{a}'&<0\,,\label{2.14}\\
\tilde{a}''&>0\,.\label{2.15}
\end{eqnarray}
Additionally, we need to impose that in each frame the Universe is
expanding, hence
\begin{eqnarray}
\tilde{a}'-\frac{1}{2}[1+f(\phi)]^{-1}f_{\phi}\phi'\tilde{a}&>0\,,\label{2.16}\\
\tilde{a}'&>0\,.\label{2.17}
\end{eqnarray}

\subsubsection{Example I}

As a first example, let us consider a de-Sitter expansion,
$\tilde{a}(\tilde{t})=\tilde{a}_{0}e^{\tilde{H}_{0}\tilde{t}}$ and
we assume that the function coupling is chosen as follows,
\begin{equation}
f(\phi )=\frac{1-\alpha \phi }{\alpha \phi }\ , \label{2.18}
\end{equation}
where $\alpha$ is a constant. As a first task, we need to
demonstrate that such kind of solution really exists. Equations
(\ref{2.5})-(\ref{2.6}) indicate that (recall that we assumed
$\kappa^2=1$),
\begin{align}
\rho_{\phi}=6\tilde{H}_0^2\,, \ \ p_{\phi}=-6\tilde{H}_0^2\ ,
\label{2.19}
\end{align}
so by using definitions (\ref{rho})-(\ref{p}) we easily find,
\begin{align}
(\phi')^2W(\phi)=0\,, \ \ U(\phi)=6\tilde{H}_0^2\ . \label{2.20}
\end{align}
Thus, we have three possibilities to satisfy these relations: either
$\phi'=0$ together with $W=0$ or $W\neq 0$ and finally if $\phi'\neq
0$ whereas $W=0$. Let us focus on the last possibility. In this case
from  Eq. (\ref{W}) we have,
\begin{align}
W(\phi)=0\,\rightarrow \ \ \omega(\phi)=\frac{-3}{\alpha\phi^3}\
\label{2.21}
\end{align}
and the potential may be found from Eq. (\ref{U}) and it reads,
\begin{align}
V(\phi)=\frac{6\tilde{H}_0^2}{\alpha^2\phi^2}\ . \label{2.22}
\end{align}
In Ref. \cite{Elizalde:2008yf} it was explicitly demonstrated that
the solution in this case is,
\begin{align}
\phi=\tilde{t}=\alpha^{\frac{-1}{3}}\left (\frac{3}{2}t \right
)^{\frac{2}{3}}\ . \label{2.23}
\end{align}
With regard to Eq. (\ref{2.17}) we have
$\tilde{a}'=\tilde{H}_0\tilde{a}>0$. The expression (\ref{2.16})
reads $\tilde{a}\left [\tilde{H}_0+(2\tilde{t})^{-1}\right ]>0$ and
it is also true for any time $\tilde{t}$. With regard to
(\ref{2.15}) we have $\tilde{a}''=\tilde{H}_0^2\tilde{a}>0$. Finally
with regard to the expression (\ref{2.14}) we have,
\begin{align}
{\frac{\tilde{a}}{\alpha\tilde{t}^3}}\left(
2\tilde{H}_0^2\tilde{t}^2 + \tilde{H}_0\tilde{t} -1 \right)<0\ ,
\label{2.24}
\end{align}
and by also taking into account that $\tilde{t}>0$, we find that the
expression (\ref{2.14}) is satisfied for
$0<\tilde{t}<1/(2\tilde{H}_0)$.

It is interesting to note, that case (\ref{2.18}) may be easy
generalized to
\begin{equation}
f(\phi )=\frac{1-\alpha \phi^n }{\alpha \phi^n }\ . \label{2.25}
\end{equation}
In this case, the general solution $\phi(\tilde{t})=\tilde{t}$ is
valid, but the time $t$-dependence, will be completely different.
The condition (\ref{2.24}) is transformed as follows,
\begin{align}
{\frac{\tilde{a}}{\alpha\tilde{t}^{n+2}}}\left(
2\tilde{H}_0^2\tilde{t}^2 + n\tilde{H}_0\tilde{t} -n \right)<0\ ,
\label{2.26}
\end{align}
and it is satisfied when the following conditions hold true,
$0<\tilde{t}<\left[-\frac{n}{4}+\frac{n}{4}\sqrt{1+\frac{8}{n}}
\right]/\tilde{H}_0$.

\subsubsection{Example II}

Now let us try to modify the previous example in such a way so that
it becomes more realistic. Firstly, notice that the solution
$\phi=\tilde{t}$ is the general solution, as it was also
demonstrated in previous works \cite{Elizalde:2008yf}. So by adding
by hand some specific cosmological behavior $\tilde{a}(\tilde{t})$
we may find the functions $\omega$ and $V$ which generate such a
cosmological evolution. In this sense, this method is analog of the
reconstruction procedure \cite{Elizalde:2008yf}. In the previous
example, the cosmological evolution was that of an exact de Sitter
solution, but it is well known that inflation is actually a quasi-de
Sitter solution. Thus consider the following quasi-de Sitter
cosmological evolution,
\begin{equation}
\tilde{H}(\tilde{t})=H_0-h(\tilde{t}) , \label{2.25}
\end{equation}
where $H_0$ is some positive constant and $h$ is a slow-varying
function of the time variable $\tilde{t}$. This function will
determine the exit from inflation, and also it crucially affects the
spectrum of cosmological perturbation, but for the moment we leave
this undetermined. Note also that the expression in Eq. (\ref{2.14})
(which was derived by using general considerations) contains only
the functions $\tilde{a}$, $\tilde{a}'$, $\tilde{a}''$ and also the
function $f$ and its derivatives. Let us assume that the function
$f$ has the form (\ref{2.25}), in effect we have,
\begin{align}
\tilde{a}' &=\tilde{a}H_0-\tilde{a}h\,,\label{2.26}\\
\tilde{a}''
&=\tilde{a}H_0^2-2\tilde{a}hH_0-\tilde{a}h^2-\tilde{a}h'.
\label{2.27}
\end{align}
By substituting these expressions in Eq. (\ref{2.14}) and by
rearranging the terms we get,
\begin{equation}
\frac{\tilde{a}}{\alpha\tilde{t}^{n+2}}\left [2\tilde{t}^2H_0^2
+nH_0\tilde{t} -n - \left ( 2\tilde{t}^2h' +2\tilde{t}^2h^2 +
4H_0h\tilde{t}^2 +nh\tilde{t} \right) \right ] <0 . \label{2.28}
\end{equation}
This is an algebraic\footnote{We suppose that $h(\tilde{t})$ is some
known function} expression of time variable $\tilde{t}$, provides us
an explicit example where acceleration (inflation) in one frame
corresponds to deceleration in another. It is quite clear that the
expression (\ref{2.28}) may be valid at least for some time
instance, but let us demonstrate this fact explicitly by using a
concrete example. We assume that $h(\tilde{t})=k\ln(1+\tilde{t})$.
For small values of time (early universe) we may decompose $h$ in
Taylor series $h\thickapprox k\tilde{t}$, $h'\thickapprox
k-k\tilde{t}$ and the expression in square brackets (it is clear
that the coefficient of it, is always positive) becomes,
\begin{equation}
2\tilde{t}^2H_0^2 +nH_0\tilde{t} -n - 2\tilde{t}^2k+ 2\tilde{t}^3k -
2\tilde{t}^4k^2 - 4H_0k\tilde{t}^3 - nk\tilde{t}^2 <0 . \label{2.29}
\end{equation}
By keeping only leading order terms, from the inequality
(\ref{2.29}), we find that (\ref{2.14}) in this case will be true
for,
\begin{equation}
0<\tilde{t} \lesssim\frac{1}{H_0} . \label{2.30}
\end{equation}
Hence we showed explicitly that accelerating expansion in one frame
corresponds to decelerating expansion in the other frame.


\section{Discussion} \label{sec:3}

The question which of the different mathematical frames is the
physical one is very profound, and in this paper we investigated how
it is possible to have accelerating expansion of the Universe in one
frame and decelerating expansion in the other. The frames which we
studied are the Jordan frame of an $F(R)$ gravity, the Einstein
frame which corresponds to a minimally coupled scalar-tensor theory,
and finally the frame which corresponds to a non-minimally coupled
scalar field. As we showed, if certain conditions are satisfied, it
is possible to have acceleration in the Jordan frame, and when the
metric is conformally transformed to the Einstein frame, the
transformed metric can describe a decelerating Universe. The same
situation can occur when one considers a non-minimally coupled
scalar field that is conformally transformed to the Einstein frame.
We illustrated our arguments by using several characteristic
examples. According to our findings, the various mathematical frames
are physically equivalent when conformal invariant quantities are
considered, like for example quantities related to the comoving
curvature. However, in principle, the physical interpretation in the
various frames can be different if non-conformal invariant
quantities are used.

Finally, let us note that the question of physical equivalence of
the different frames becomes much more involved when quantum effects
are taken into account, see for example Ref. \cite{last1} for
minimal and non-minimal frames, and also consult \cite{last2}.

\begin{acknowledgments}
 SB is supported by the Comisi{\'o}n Nacional de Investigaci{\'o}n Cient{\'{\i}}fica y Tecnol{\'o}gica (Becas Chile Grant No.~72150066).
 This work was partially supported by the Russian Science Foundation (RSF) grant 16-12-10401 (P.V.T.), and by Min. of Education and Science of Russia (S.D.O
and V.K.O). Also this work is supported by MINECO (Spain), project
 FIS2013-44881 (S.D.O)
\end{acknowledgments}

\end{document}